\documentclass[12pt]{article}
\usepackage{graphicx}

\newcommand\pubnumber{SLAC--PUB--15178}
\newcommand\pubdate{v3: March 2013}

\def\SLAC{SLAC,
    Stanford University, Menlo Park, California 94025 USA}
\def\doeack{\footnote{Work supported by the US Department of Energy,
                     contract DE--AC02--76SF00515.}}

\def\Title#1{\begin{center} {\Large #1 } \end{center}}
\def\Author#1{\begin{center}{ \sc #1} \end{center}}
\def\Address#1{\begin{center}{ \it #1} \end{center}}

\newcommand\pubblock{\rightline{\begin{tabular}{l} \pubnumber\\
         \pubdate \end{tabular}}}
\newenvironment{Abstract}{\begin{quotation} \begin{center}
                       ABSTRACT
     \end{center}\bigskip  }{\end{quotation}}

\def\Acknowledgements{\bigskip  \bigskip \begin{center} \begin{large}
             \bf ACKNOWLEDGEMENTS \end{large}\end{center}}

\def\beq{\begin{equation}}
\def\eeq#1{\label{#1}\end{equation}}
\def\eeqn{\end{equation}}

\newenvironment{Eqnarray}%
   {\arraycolsep 0.14em\begin{eqnarray}}{\end{eqnarray}}
\def\beqa{\begin{Eqnarray}}
\def\eeqa#1{\label{#1}\end{Eqnarray}}
\def\eeqan{\end{Eqnarray}}

\def\leqn#1{(\ref{#1})}

\let\bar=\overbar

\def\lsim{\mathrel{\raise.3ex\hbox{$<$\kern-.75em\lower1ex\hbox{$\sim$}}}}
\def\gsim{\mathrel{\raise.3ex\hbox{$>$\kern-.75em\lower1ex\hbox{$\sim$}}}}

\def\del{\partial}
\def\Dslash{\not{\hbox{\kern-4pt $D$}}}
\def\dslash{\not{\hbox{\kern-2pt $\del$}}}

\def\ee{e^+e^-}

\def\mz{m_Z}

\def\mw{m_W}

\def\msb{{\bar{\scriptsize M \kern -1pt S}}}

\def\drb{{\bar{\scriptsize D \kern -1pt R}}}

\makeatletter
\def\section{\@startsection{section}{0}{\z@}{5.5ex plus .5ex minus
 1.5ex}{2.3ex plus .2ex}{\large\bf}}
\def\subsection{\@startsection{subsection}{1}{\z@}{3.5ex plus .5ex minus
 1.5ex}{1.3ex plus .2ex}{\normalsize\bf}}
\def\subsubsection{\@startsection{subsubsection}{2}{\z@}{-3.5ex plus
-1ex minus  -.2ex}{2.3ex plus .2ex}{\normalsize\sl}}

\renewcommand{\@makecaption}[2]{%
   \vskip 10pt
   \setbox\@tempboxa\hbox{\small #1: #2}
   \ifdim \wd\@tempboxa >\hsize     
       \small #1: #2\par          
     \else                        
       \hbox to\hsize{\hfil\box\@tempboxa\hfil}
   \fi}

 \def\citenum#1{{\def\@cite##1##2{##1}\cite{#1}}}
 
\newcount\@tempcntc
\def\@citex[#1]#2{\if@filesw\immediate\write\@auxout{\string\citation{#2}}\fi
  \@tempcnta\z@\@tempcntb\m@ne\def\@citea{}\@cite{\@for\@citeb:=#2\do
    {\@ifundefined
       {b@\@citeb}{\@citeo\@tempcntb\m@ne\@citea\def\@citea{,}{\bf ?}\@warning
       {Citation `\@citeb' on page \thepage \space undefined}}%
    {\setbox\z@\hbox{\global\@tempcntc0\csname b@\@citeb\endcsname\relax}%
     \ifnum\@tempcntc=\z@ \@citeo\@tempcntb\m@ne
       \@citea\def\@citea{,}\hbox{\csname b@\@citeb\endcsname}%
     \else
      \advance\@tempcntb\@ne
      \ifnum\@tempcntb=\@tempcntc
      \else\advance\@tempcntb\m@ne\@citeo
      \@tempcnta\@tempcntc\@tempcntb\@tempcntc\fi\fi}}\@citeo}{#1}}
\def\@citeo{\ifnum\@tempcnta>\@tempcntb\else\@citea\def\@citea{,}%
  \ifnum\@tempcnta=\@tempcntb\the\@tempcnta\else
  {\advance\@tempcnta\@ne\ifnum\@tempcnta=\@tempcntb \else\def\@citea{--}\fi
    \advance\@tempcnta\m@ne\the\@tempcnta\@citea\the\@tempcntb}\fi\fi}
\makeatother

\begin{document}
\begin{titlepage}
\pubblock

\vfill
\Title{Comparison of LHC and ILC Capabilities for Higgs Boson Coupling Measurements}
\vfill
\Author{Michael E. Peskin\doeack}
\Address{\SLAC}
\vfill
\begin{Abstract}
I estimate the accuracies on Higgs boson coupling constants that
experiments at the Large
Hadron Collider and the International Linear Collider are 
capable of reaching over the long term.
\end{Abstract}
\vfill
\vfill
\newpage
\tableofcontents
\end{titlepage}

\def\thefootnote{\fnsymbol{footnote}}
\setcounter{footnote}{0}

\section{Introduction}

Now that a scalar boson of mass about 125~GeV has been discovered by
the ATLAS and CMS 
experiments~\cite{Higgsseminar,ATLASHiggs,CMSHiggs}, the
question of the hour is:  Is this the Higgs boson?   The key property
of the Standard Model Higgs boson is that its coupling to each fermion
and boson species are proportional to its mass.  For a boson at
125~GeV, we can test this for a large number of Standard Model
species.

How accurate must these tests be?    Today, we would be pleased to
achieve accuracies of 30-50\% in the couplings.   Agreement to this
accuracy would make a strong case that the particle discovered at the
LHC is indeed the Higgs boson.  

However, there is a second interesting question that should be
addressed.  Many models with new physics beyond the Standard Model
contain a light Higgs boson with properties very similar to the Higgs
boson of the Standard Model.   Haber has called attention to the 
`Decoupling Limit'
as a generic phenomenon in new physics models.  
In a large region of parameter space, the Higgs couplings in such models
differ from those in the Standard Model by 10\% or less~\cite{Haber}. 
 Recently,
Gupta,
Rzehak, and Wells have illustrated this phenomenon in a wide variety
of
 models~\cite{GRW}. 

As we look to the future in particle physics, it is a very interesting
question how capably planned and proposed experiments
 will be able to probe the couplings
of the Higgs boson at such a high level of precision.    In this note,
I will estimate the accuracies that can be achieved by experiments at 
Large Hadron Collider and at the International Linear Colllider
 over the long term.  It is challenging to predict the size of the
error bars for experiments that will be carried out many years from
now.
I hope that this note will at least provide a plausible 
methodology, leading to estimates whose quality is
straightforward to evaluate.  Ultimately, the ATLAS and CMS experiments 
will need to make more definitive estimates of their ultimate
sensitivities. But these are not yet available.   This paper represents
my attempt to fill this gap.

There have been many previous attempts to estimate the ultimate 
sensitivity of the LHC to the Higgs boson couplings.  These include
works of Zeppenfeld {\it et al}~\cite{Zepp}, Belyaev and Reina~\cite{Belyaev},
and D\"uhrssen {\it et al}
\cite{Duhrssen,Lafaye,Klute,SFitter}.  I regard the work of D\"uhrssen, Plehn, and
collaborators as particularly important, and I will borrow many ideas 
from this work in the following.  In general, the method here will be more
simplistic but, I hope, more transparent than that used in \cite{Lafaye}.
and \cite{Klute}. 

This version 3 of the paper takes account of new analysis by ATLAS
and CMS~\cite{ATLASESS,CMSESS,newCMS} and uses the final estimates of the accuracy of 
 ILC Higgs measurements published in the ILC Technical Design
 Report~\cite{ILCDBD}.

\section{ Methodology}

In this  paper, I will always consider the couplings of a specific CP even
scalar state to particle-antiparticle states.   This state
should be a scalar particle at a fixed mass.  From here on, I will
refer to this particle without further apology 
as `the Higgs boson'.  I will write the Higgs
coupling to a particle $A$ and its antiparticle as $g(hAA)$.
This coupling constant will be
associated with a Lorentz structure  in a canonical way that depends
on the spin of the particle $A$.  I will work directly in terms of
the Higgs couplings
$g(hAA)$; the Higgs boson partial widths to $A\bar A$ are
proportional to the squares of these  quantities.  

We should treat these Higgs boson couplings as
completely unknown,  
 to be  determined by experimental measurements.  
Both for the LHC and for the ILC, the range of observables that will
eventually be measured is broad enough to permit model-independent
fits of to all couplings independently.   Note that I will not assume
any {\it a priori} relation between the tree-level couplings of the
Higgs boson to fermions and massive vector bosons and the loop-level
couplings to $gg$ and $\gamma\gamma$.  In the Standard Model, the 
loop-level couplings are related to the Higgs couplings to $t$ and
$W$; however, in a more general model, other heavy particles can
contribute to the loop diagrams.   A model-independent approach should
fit the values of the loop diagrams separately from the direct
couplings to $t$ and $W$ final states.

It is completely straightforward to measure Higgs boson couplings in a
model-independent way at the ILC.
  The ILC will make it possible to measure the cross
section $\sigma(\ee\to Z^0h^0)$ without  reference to branching ratios
of the Higgs by observing the recoil  $Z^0$ at a fixed lab energy.
Individual branching ratios can then be measured directly as the fractions
of this total cross section in which the specific final state is observed.

For the LHC, it is less obvious that such a general analysis can be
performed.  However, it is possible to make fits to Higgs couplings
that are  almost model-independent
through the use of a simple and very weak theoretical assumption.
In the Lagrangian for $SU(2)\times U(1)$ gauge fields coupled to an
arbitrary number of Higgs fields in arbitrary representations, it is
always true that the various Higgs fields with vacuum expectation
values make positive contributions to the $W$ and $Z$ masses.   The
Higgs couplings to $WW$ and $ZZ$ arise by differentiating these
contributions with respect to the Higgs field vev, so it makes sense
that these couplings are also positive terms whose sum is
set
by the $W$ and $Z$ masses.   The precise version of this statement,
derived by Gunion, Haber, and Wudka ~\cite{GHW}, is that, in a model
with a CP-conserving Higgs sector in which only CP = +1 fields have
nonzero vevs and in which couplings of doubly charged Higgs fields
$g(W^+W^+\phi^{--})$ are absent,
\beq
      \sum_k  g^2(\phi^0_k WW)^2/g^2 = ( 4 \mw^2 - 3 \cos^2\theta_w \mz^2)
\eeq{sumrule}
To high accuracy, the right-hand side can be replaced by $\mw^2$.
Then it follows that, for any individual neutral Higgs boson state
$h^0$, 
\beq
         |g(hWW) |   <    g(hWW)|_{SM} \ ,
\eeq{bound}
where the right-hand side is the value of the $hWW$ coupling in the
Standard Model. 
Similarly, 
\beq
         |g(hZZ) |   <    g(hZZ)|_{SM} \ ,
\eeq{boundZ}
  The importance of this constraint was recognized and
first applied to the interpretation of LHC Higgs observables  by
D\"uhrssen {\it et al}~\cite{Duhrssen}. I follow their logic in 
the discussion below. 

In some analyses, the
constraints \leqn{bound} and \leqn{boundZ} 
are applied together with the constraint
\beq
         g(hWW)/g(hZZ)   =  \cos^2\theta_w 
\eeq{boundrelation}
which is valid in models in which the Higgs boson is a linear
combination of $SU(2)$ singlets and doublets only.  I do not apply that
constrain here.  The measurement of the ratio of the $W$ and $Z$ couplings
is an important basic test of the nature of the Higgs boson and needs to be 
carried out however plausible the relation \leqn{boundrelation} might be.

Higgs boson observables at the LHC are either 
ratios of branching ratios or measured rates, proportional to cross
sections times branching ratios.    In the former case, the overall
scale of the branching ratios cancels out.  In the latter case, the
quantity measured, for the observable $\sigma(A\bar A\to h) BR(h\to
B\bar B)$,  is proportional to 
\beq
                {   g^2(hAA)\, g^2(hBB)\over \Gamma_T} \ ,
\eeq{grat}
where $\Gamma_T$ is the total width of the Higgs.   To determine
the absolute magnitudes of the Higgs couplings, we must have some
information about this total width.
For a Standard Model Higgs boson at 125~GeV, the predicted width is
4~MeV.  So the Higgs boson width is not expected to be directly
measurable at any collider.   
 
It is possible, though, to constrain the total width of the Higgs boson 
from the measurement of $\sigma\cdot BR$ observables at the LHC.  Consider,
for example, the measurement of the rate for $WW$ fusion production of a 
Higgs boson which then decays to $WW^*$.  Writing 
\beq
        \Gamma_T = \Gamma(h\to WW^*)/BR(h\to WW^*)\ , 
\eeq{GammaTtoBR}
we see that the rate is proportional to 
\beq
                (g^2(hWW))\, BR(h\to WW^*) \ .
\eeq{gGam}
The Higgs branching ratio to $WW^*$ must be less than 1, so we obtain a 
lower bound on $g^2(hWW)$ and on $\Gamma_T$.  We can improve this lower
bound by adding in the branching ratios to other Higgs decay modes observed
at the LHC, determined relative to $BR(h\to WW^*)$ 
from measurements of ratios of branching ratios.  In fact, it is possible to 
observe at the LHC almost all of the significant decay modes of the 
Standard Model Higgs boson.  Only for $h\to c\bar c$, a mode with a 
3\% branching ratio in the Standard Model, is there currently no strategy 
for observation.  The decay $h\to gg$ is not directly 
observable, but the coupling
of the Higgs boson to $gg$ enters the analysis through the cross section 
for Higgs production from gluon fusion.  So this lower bound on $\Gamma_T$ 
could in principle be pushed up to 
97\% of the Standard Model value.  

Still, there could in principle 
be other decay modes involving particles outside the Standard Model that 
are not observable at the LHC and could raise the value of $\Gamma_T$.
Thus, to complete the analysis, we also need an upper bound on $\Gamma_T$. 
This is provided by the inequalities \leqn{bound} and \leqn{boundZ}.
In the Standard Model, the theoretical values of the 
upper and lower bounds are within 3\% of 
one another and thus constrain the Higgs
width $\Gamma_T$ to an accuracy greater than the accuracy of the actual
measurements that will be made.
 
We then proceed in the following way:  Write the deviations from the 
Higgs couplings as
\beq
      {  g(hAA)\over g(hAA)|_{SM}} = 1 + d(A)
\eeq{ddefin}
 I will include only one possible decay channel not 
included in the Standard Model, a decay to invisible decay modes, defining
$d$ for that channel by
\beq
      d^2(\mbox{inv}) = BR(h\to \mbox{inv})\ .
\eeq{dinvis}
The invisible mode of Higgs decay can be observed at the LHC using the
vector boson fusion process~\cite{Eboli}.  It is possible that there are
additional non-Standard modes of Higgs decay that are not visible at 
the LHC.  My fit includes the $c\bar c$ mode of Higgs decay, which is not
visible at the LHC; other possible non-visible modes are taken into 
account here.

I will take the variables $d(A)$, with flat priors, as the basic variables
for this analysis.

In terms of $d(A)$, deviations in the cross section
are given by 
\beq
       {\sigma(A\bar A \to h)\over \sigma(A\bar A\to h)|_{SM}} = 
               (1 + d(A))^2 \ .
\eeq{devsigma}
Deviations in ratios of branching ratios are given by 
\beq
      { BR(h \to A\bar A )/BR(h \to B\bar B)\over 
         BR(h \to A\bar A )/BR(h \to B\bar B)|_{SM}} = 
              { (1 + d(A))^2 \over (1 + d(B))^2 } \ .
\eeq{devBRrat}
Deviations in rates are given by 
\beq
 {\sigma(A\bar A \to h)BR(h\to B\bar B)\over \sigma(A\bar A\to h)
 BR(h\to B\bar B)|_{SM}} = 
              { (1 + d(A))^2 (1 + d(B))^2\over D\Gamma }
\eeq{devsigmaBR}
where 
\beq
   D\Gamma = 
      (\sum_X BR(h\to X\bar X)|_{SM} \cdot (1+d(X))^2 )/(1.0 - d^2(\mbox{inv}) )\ , 
\eeq{devsigmaBRGam}
where the expression on the right contains the Standard Model branching
fractions for a Higgs boson of mass 125~GeV.s  For the special case of
invisible
decays, 
\beq
 {\sigma(A\bar A \to h)BR(h\to \mbox{invis})\over \sigma(A\bar A\to h)|_{SM}} = 
       (1 + d(A))^2 ( d(\mbox{invis}))^2  \ .
\eeq{devsigmaBinv}

To estimate errors on the parameters $d(A)$, I will work from a list 
of measurements that can be made at the LHC and the ILC.   I will assume, 
somewhat ideally, that each measurement has the Standard Model value as 
its outcome and that the probability distribution for deviations from 
that outcome is Gaussian.   This produces a likelihood function
\beq
      {\cal L} = \prod_i exp[- {\cal D}_i^2/2 \sigma_i^2]  \cdot  {\cal C} \ ,
\eeq{likeli}
where the variables ${\cal D}_i$ are combinations of the form
 \leqn{devsigma}, 
\leqn{devBRrat}, or \leqn{devsigmaBR} above, and ${\cal C}$ is a product 
of theta functions implementing the constraints \leqn{bound} and \leqn{boundZ}
and the constraints that $(1 + d(A)) > 0$ for all $A$ 
(and $d(\mbox{inv}) > 0$).  I integrated this likelihood function using 
VEGAS, formed the probability distribution for each variable, and computed
for each the boundaries of the 68\% confidence interval about the mean. 
These error intervals are tabulated for the various scenarios in 
Table~\ref{tab:results} below. This approach can be described as
`naive Bayesian'.  The results depend on the choice
of a flat prior for the $d(A)$; however, to the extent that the boundaries
of the confidence
intervals on the $d(A)$ are close to 0, the results become 
independent of this choice.  

In each scenario considered, the suite of
measurements does produce for each variable $d(A)$
a smooth probability distribution that decreases
monotonically from a maximum close to 0.  I see none of the pathologies
described for these probability distributions for the more complex 
likelihood function studied in \cite{Lafaye}.

\section{ Inputs}

In the previous section, I have explained a method that leads from 
input data in the form of a list of observables and their estimated
relative errors to a set of confidence intervals for the variables
$d(A)$.   We must now discuss what input data should be used.

  On the LHC side, we would like
to include the ultimate, systematics-limited errors on the
measurements of Higgs observables.  Unfortunately, there is no
comprehensive,
up-to-date study that reflects the current understanding of Higgs
measurements or the  measured capabilities of the ATLAS and CMS
detectors.  The  most recent comprehensive study of the ultimate 
LHC capabilities
for Higgs measurements
is the 2003 Ph.D. thesis of  Michael
D\"uhrssen~\cite{Dthesis}.  This thesis estimated the expected errors
on measurements of Higgs observables by the ATLAS detector for 
300~fb$^{-1}$ of  data.  By that point, these measurements will be
systematics-limited.   Very recently, the ATLAS and CMS collaborations
have reported updated estimates for a subset of their possible Higgs
boson measurements in their submissions to the European Strategy
Study~\cite{ATLASESS,CMSESS}.
  In Table~\ref{tab:inputsLHC}, I have attempted to compile the 
conclusions of these reports for the achievable accuracies on the
 important Higgs observables for
a Higgs boson of mass 125~GeV.   This table has changed from version 1
to version 2 of this paper to reflect the numbers given in
\cite{ATLASESS,CMSESS}.
For an explanation of the changes, see the Appendix.

Measurements of $\sigma\cdot BR$ are
also subject to significant theoretical systematic errors from the QCD
calculations of the cross section.  For specific measurements such as
those for the final state $WW$, which requires jet vetos, and boosted
Higgs measurements, which require a Higgs at high $p_T$, the selection
of events lowers the
precision of the QCD cross section by one order in $\alpha_s$.
I have guessed at these errors in
the Table and added the experimental and theoretical 
errors in quadrature.  I have assumed that the
various Higgs production processes can be cleanly separated, with no
correlations in the measurement errors.  I have also ignored 
correlated theoretical errors in the estimation of common production 
cross sections.  These two assumptions work to reduce the estimated
errors resulting from the LHC measurements.

D\"uhrssen's thesis must be used with some caution.  First,
D\"uhrssen's thesis ignores complications from pileup.   But, more
importantly, his work assumed
that the processes $pp\to W,Z + h$ and $pp\to t\bar t h$, with $h \to
b\bar b$  would be
straightforwardly observable.  The small errors assumed for these
processes played an important role in the optimistic conclusions of
the fit \cite{Duhrssen}. The coupling $g(hb\bar b)$ plays a central
role in the overall analysis, since the mode $h\to b\bar b$ accounts for
60\% of the total with of a 125~GeV Higgs boson in the Standard Model.  
The coupling $g(hbb)$
is constrained only by the these two processes, and so the accuracy
of $\sigma\cdot BR$ measurements for this processes have a crucial
effect on the final results.
  In the mid-2000's, studies by the ATLAS and
CMS experimental groups gave very pessimistic conclusions about the
visibility of of these modes.   More recently, there is optimism
again, due to the development of methods for tagging boosted objects
in the LHC environment~\cite{Butterworth,Plehn}.   However, this
technique is still unproven, and it is not well understood how to
estimate the efficiency of the event selection for the purpose of
measuring a $\sigma\cdot BR$.   In the Table, I have assigned these
two processes a 25\% experimental systematic error, which I believe is
optimistic.

For the direct measurement of invisible decays of the Higgs, I
estimate an experimental error of 20\%, plus a theoretical systematic
error estimated by Bai, Draper, and Shelton~\cite{BDS} to be 24\%.

Persuasive estimates of the ultimate errors of the LHC measurements
can come only from the ATLAS and CMS collaborations in analyses that 
reflect their best
understanding of the capabilities of their detectors based on 
the detector performance in the 7 and 8~TeV runs. ATLAS and CMS 
have begun that work 
in \cite{ATLASESS,CMSESS,newCMS}, and these papers represent their
 best estimates at the moment.

 The 
estimates given here reflect 1 LHC detector accumulating 300~fb$^{-1}$
of data.  I do not consider the problem of combining results from two
detectors in the presence of dominant systematic errors.  In 
\cite{ATLASESS,CMSESS,newCMS}, ATLAS and CMS also estimated errors for
3000~fb$^{-1}$ of data from the High-Luminosity LHC.  Such estimates,
though, cannot be done as a straightforward extrapolation of current
performance.  It is probably true that some improvements can be made,
but this is not obvious given the very difficult experimental
conditions of the HL-LHC.  The ATLAS and CMS extrapolations are 
discussed further in the Appendix.

\begin{figure}[p]
\begin{center}
\includegraphics[width=6.0in]{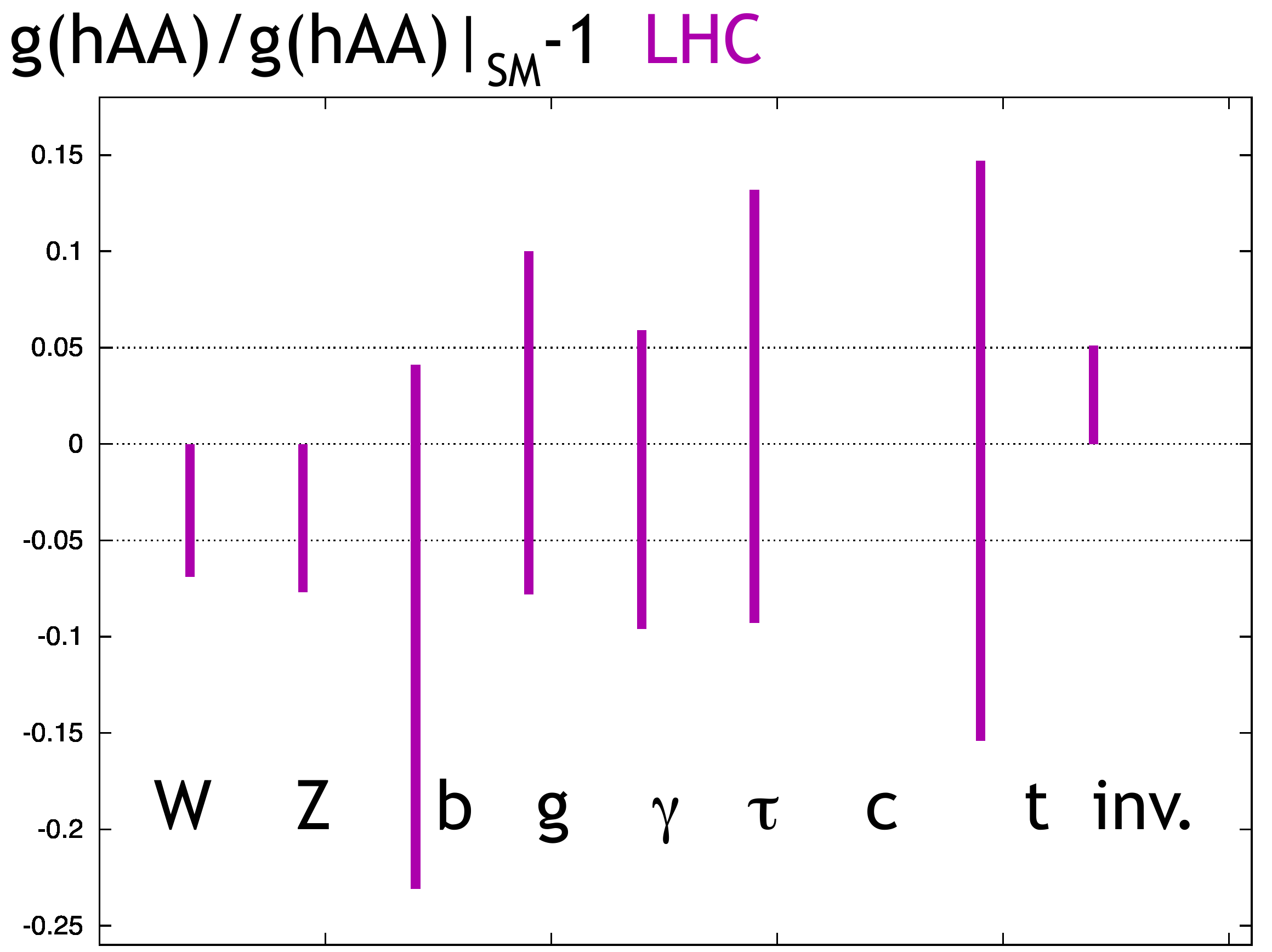}
\caption{Capabilities of LHC for
  model-independent measurements of Higgs boson couplings.   The plot
  shows the  1~$\sigma$ confidence intervals for LHC at 14~TeV with
  300 fb$^{-1}$ as they emerge from my fit.  Deviation of the central
  values from zero indicates a bias, which can be corrected for.  The 
upper limit on the $WW$ and $ZZ$ couplings arises from the constraints
\leqn{bound} and \leqn{boundZ}.
   No error is estimated for $g(hcc)$.  The bar for the invisible
   channel gives the 1 $\sigma$ upper limit on the {\it branching
     ratio}. The marked horizontal band 
 represents a 5\% deviation from the Standard Model prediction for the
     coupling.}
\label{fig:resultsLHC}
\end{center}
\end{figure}

\begin{figure}[p]
\begin{center}
\includegraphics[width=6.0in]{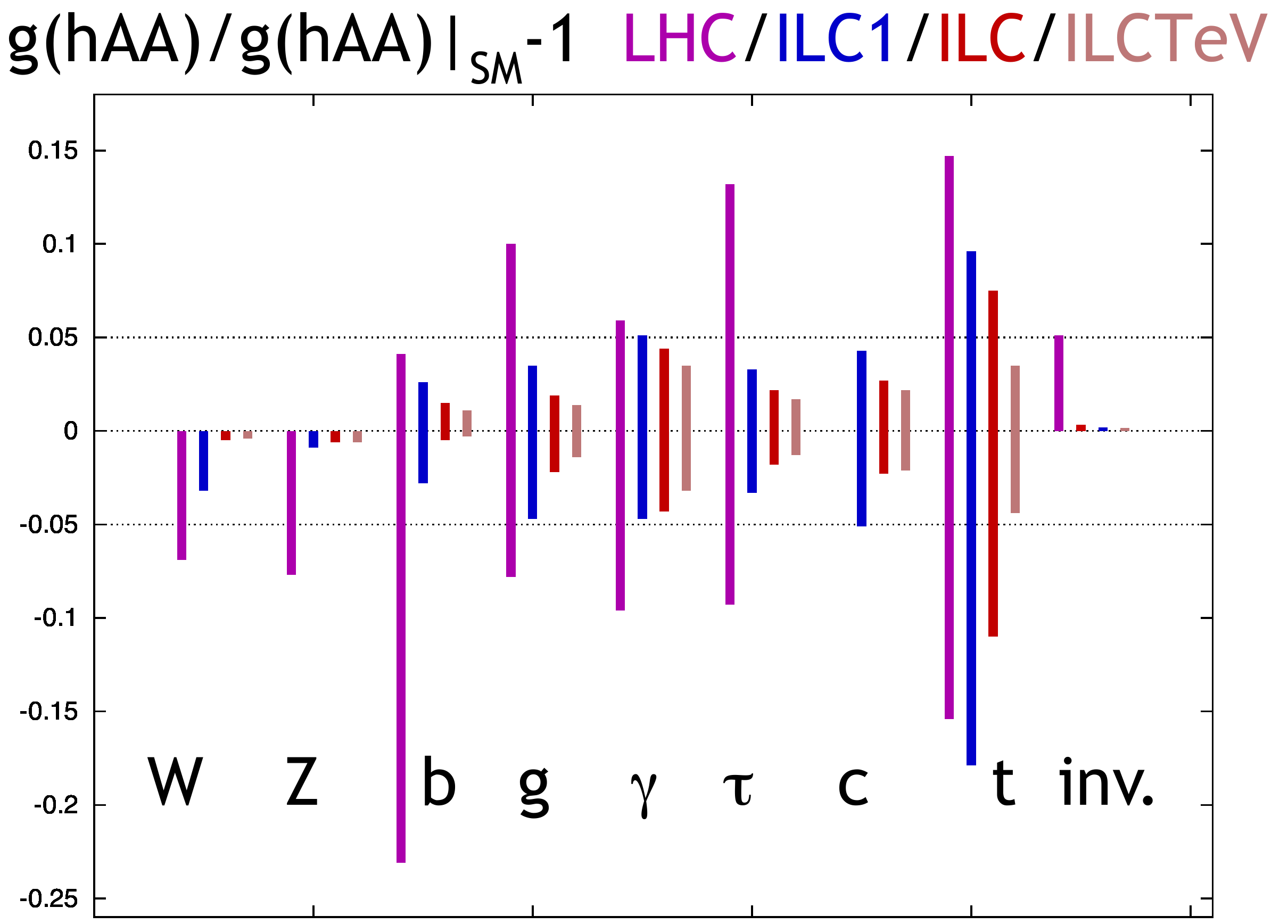}
\caption{Comparison of the capabilities of LHC and ILC for
  model-independent measurements of Higgs boson couplings.   The plot
  shows (from left to right in each set of error bars) 
1~$\sigma$ confidence intervals for LHC at 14~TeV with 300 fb$^{-1}$, 
     for
  ILC at 250~GeV and 250~fb$^{-1}$ (`ILC1'), for  the full ILC program up
  to 500~GeV with 500~fb$^{-1}$ (`ILC'), 
 and for a program with 1000~fb$^{-1}$ for 
   an upgraded ILC at 1~TeV (`ILCTeV').  More details of the
   presentation are given in the caption of Fig.~\ref{fig:resultsLHC}.  The marked horizontal band 
 represents a 5\% deviation from the Standard Model prediction for the
     coupling.} 
\label{fig:resultsILC}
\end{center}
\end{figure}
\begin{table}[p]
\begin{center}
\begin{tabular}{lc} 
Observable   &    Expected Error  (experiment $\oplus$ theory)\\
\hline
LHC at 14 TeV with 300 fb$^{-1}$  \\ 
\hline
$\sigma(gg)\cdot BR(\gamma\gamma)$ &    0.06 $\oplus$  0.13  \\
$\sigma(WW)\cdot BR(\gamma\gamma)$ &   0.15 $\oplus$  0.10 \\
$\sigma(gg)\cdot BR(ZZ)$ &                     0.08 $\oplus$  0.08        \\ 
$\sigma(gg)\cdot BR(WW)$ &                     0.09 $\oplus$  0.11        \\ 
$\sigma(WW)\cdot BR(WW)$ &                   0.27 $\oplus $ 0.10  \\ 
$\sigma(gg)\cdot BR(\tau^+\tau^-)$ &  0.11 $\oplus $  0.13 \\ 
$\sigma(WW)\cdot BR(\tau^+\tau^-)$ &  0.15 $\oplus $  0.10 \\ 
$\sigma(Wh)\cdot BR(b\bar b) $      &           0.25 $\oplus$ 0.20\\ 
$\sigma(Wh)\cdot BR(\gamma\gamma)$  &     0.24 $\oplus$ 0.10       \\ 
$\sigma(Zh)\cdot BR(b\bar b) $      &           0.25 $\oplus$ 0.20\\ 
$\sigma(Zh)\cdot BR(\gamma\gamma)$  &     0.24 $\oplus$ 0.10       \\ 
$\sigma(t\bar t h) \cdot BR(b\bar b)$  &          0.25  $\oplus$  0.20 \\ 
$\sigma(t\bar t h) \cdot BR(\gamma\gamma)$  &   0.42$\oplus$  0.10\\
$\sigma(WW)\cdot BR(\mbox{invisible})$ &   0.2 $\oplus$ 0.24\\  \hline
\end{tabular}
\caption{Input data for the fits to Higgs couplings from LHC
  measurements.  See the discussion in the Appendix.}
\label{tab:inputsLHC}
\end{center}
\end{table}
\begin{table}[p]
\begin{center}
\begin{tabular}{ll} 
Observable   &    Expected Error \\
\hline
ILC at 250 GeV with 250 fb$^{-1}$ \\
\hline
$\sigma(Zh)$  &   0.025  \\ 
$\sigma(Zh)\cdot BR(b\bar b) $ &   0.011 \\ 
$\sigma(Zh)\cdot BR(c\bar c) $ &  0.074 \\ 
$\sigma(Zh)\cdot BR(gg) $ &        0.091 \\ 
$\sigma(Zh)\cdot BR(WW) $ &       0.064\\ 
$\sigma(Zh)\cdot BR(ZZ) $ &       0.19   \\ 
$\sigma(Zh)\cdot BR(\tau^+\tau^-) $ &   0.042\\ 
$\sigma(Zh)\cdot BR(\gamma\gamma) $ &    0.38 \\  
$\sigma(WW)\cdot BR(b\bar b) $ &  0.105 \\
$\sigma(Zh)\cdot BR(\mbox{invisible})$ &   0.005 \\  \hline
ILC at 500 GeV with 500 fb$^{-1}$ \\  
\hline
$\sigma(Zh)\cdot BR(b\bar b) $ &   0.018 \\ 
$\sigma(Zh)\cdot BR(c\bar c) $ &  0.12 \\ 
$\sigma(Zh)\cdot BR(gg) $ &        0.14 \\ 
$\sigma(Zh)\cdot BR(WW) $ &       0.092   \\ 
$\sigma(Zh)\cdot BR(ZZ) $ &       0.25   \\ 
$\sigma(Zh)\cdot BR(\tau^+\tau^-) $ &   0.054\\ 
$\sigma(Zh)\cdot BR(\gamma\gamma) $ &    0.38 \\  
$\sigma(WW)\cdot BR(b\bar b) $ &   0.0066 \\ 
$\sigma(WW)\cdot BR(c\bar c) $ &  0.062 \\ 
$\sigma(WW)\cdot BR(gg) $ &        0.041 \\ 
$\sigma(WW)\cdot BR(WW) $ &       0.026   \\ 
$\sigma(WW)\cdot BR(ZZ) $ &       0.082   \\ 
$\sigma(WW)\cdot BR(\tau^+\tau^-) $ &   0.14\\ 
$\sigma(WW)\cdot BR(\gamma\gamma) $ &    0.26 \\  
$\sigma(t\bar t h)\cdot BR(b\bar b) $ &  0.25 \\  \hline
 ILC at 1 TeV with 1000 fb$^{-1}$ \\
\hline
$\sigma(WW)\cdot BR(b\bar b) $ &   0.0047 \\ 
$\sigma(WW)\cdot BR(c\bar c) $ &  0.076 \\ 
$\sigma(WW)\cdot BR(gg) $ &        0.031 \\ 
$\sigma(WW)\cdot BR(WW) $ &       0.033   \\ 
$\sigma(WW)\cdot BR(ZZ) $ &       0.044   \\ 
$\sigma(WW)\cdot BR(\tau^+\tau^-) $ &   0.035\\ 
$\sigma(WW)\cdot BR(\gamma\gamma) $ &    0.10 \\  
$\sigma(t\bar t h)\cdot BR(b\bar b) $ &  0.087 \\ 
\end{tabular}
\caption{Input data for the fits to Higgs couplings from ILC
  measurements, from \cite{ILCDBD}. }
\label{tab:inputsILC}
\end{center}
\end{table}
\begin{table}[p]
\begin{center}
\begin{tabular}{lc} 
Program   &   $1\sigma$ Confidence interval for $d(X)$  \\
\hline
LHC at 14 TeV with 300 fb$^{-1}$ &      \\ 
\hline
$h\to WW$    &  ( -0.069 , 0.000)   \\
$h\to ZZ$    &    ( -0.077 , 0.000) \\
$h \to b\bar b$  &    ( -0.231 , 0.041 )   \\
$h \to g g $ &      ( -0.078 , 0.10 )    \\
$h \to \gamma\gamma $ &       ( -0.096 , 0.059 )     \\
$h \to \tau^+\tau^-$   &    ( -0.093, 0.132 ) \\
$h \to c \bar c $  &        ---       \\ 
$h \to t\bar t $ &     ( -0.154 , 0.147 )  \\
$h \to$ invisible &   ( -0.000 , 0.226 )   \\
\hline
ILC at 250 GeV with 250 fb$^{-1}$ &   \\
\hline
$h\to WW$    &    ( -0.032 , 0.000 )     \\
$h\to ZZ$    &      ( -0.009 , 0.000 ) \\
$h \to b\bar b$  &     ( -0.028 , 0.026 )  \\
$h \to g g $ &        ( -0.047 , 0.035 )   \\
$h \to \gamma\gamma $ &     ( -0.047 , 0.051 )       \\
$h \to \tau^+\tau^-$   &    ( -0.033 , 0.033 )    \\
$h \to c \bar c $  &       ( -0.051 , 0.043 )    \\ 
$h \to t\bar t $ &    ( -0.179 , 0.096 ) \\
$h \to$ invisible &      ( -0.000 , 0.051 )  \\
\hline
ILC at 500 GeV with 500 fb$^{-1}$   &   \\
\hline
$h\to WW$    &    ( -0.005 , 0.000 )  \\
$h\to ZZ$    &      ( -0.006 , 0.000 ) \\
$h \to b\bar b$  &    ( -0.005 , 0.015 )    \\
$h \to g g $ &     ( -0.022 , 0.019 )         \\
$h \to \gamma\gamma $ &       ( -0.043 , 0.044 )     \\
$h \to \tau^+\tau^-$   &   ( -0.018 , 0.022 )  \\
$h \to c \bar c $  &      ( -0.023 , 0.027 )     \\ 
$h \to t\bar t $ &     ( -0.11 , 0.075 )  \\
$h \to$ invisible &       ( -0.000 , 0.042 ) \\
\hline
ILC at 1000 GeV with 1000 fb$^{-1}$   &  \\
\hline
$h\to WW$    &   ( -0.004 , 0.000 )   \\
$h\to ZZ$    &     ( -0.006 , 0.000 )  \\
$h \to b\bar b$  &    ( -0.003 , 0.011 )    \\
$h \to g g $ &     ( -0.014 , 0.014 )      \\
$h \to \gamma\gamma $ &     ( -0.032 , 0.035 )      \\
$h \to \tau^+\tau^-$   &    ( -0.013 , 0.017 )  \\
$h \to c \bar c $  &      ( -0.021 , 0.022 )    \\ 
$h \to t\bar t $ &         ( -0.044 , 0.035 )      \\
$h \to$ invisible &      ( -0.000 , 0.039 )  \\
\end{tabular}
\caption{Results of the fits to Higgs couplings expressed as 
1~$\sigma$ confidence intervals on $d(X)$.}
\label{tab:results}
\end{center}
\end{table}

For the ILC,  error estimates for the $\sigma\cdot BR$ are 
reported in Table~\ref{tab:inputsILC}.  I assume three stages of ILC
operation: (1) an initial stage at 250~GeV near the maximum of the 
$\ee\to hZ$ cross section, with 250~fb$^{-1}$ of data; (2) a stage at
the ILC top energy of 500~GeV, with 500~fb$^{-1}$ of data; (3) a stage
with an upgraded ILC at 1~TeV with 1000~fb$^{-1}$ of data.  The 
nominal ILC program consists of stages 1 and 2 only. The errors quoted 
for each measurement are taken from the Physics Volume of the ILC 
Tehcnical Design Report~\cite{ILCDBD}. 
These numbers are based on full-simulation results for a Standard
Model
Higgs boson of mass 120~GeV and extrapolated to a Higgs boson 
mass of 125~GeV.    The selection of $\gamma\gamma$ events in 
the ILC studies is manifestly not optimized, but I have used the
upper limit of the error range given in \cite{ILCDBD}. 
I have ignored theoretical errors on the ILC measurements, since these 
are generally at the parts per mil level.  In fact, all errors
reported are dominated by statistical errors.   If the ILC performs
better 
than the current conservative projections, these errors will be improved.

The 1~TeV ILC program will have other results that are interesting for
Higgs physics.  These include measurements of $g(h\mu\mu)$ at the 16\%
level  and of the 
Higgs self-coupling at the 20\% level, according to the 
current best understanding~\cite{ILCDBD}.  Linear Collider
measurements of the Higgs boson at higher energy are discussed in the
CLIC Conceptual Design Report~\cite{CLICCDR}.

The fits to ILC data include the data from LHC.   The fit to ILC data
from the full program includes the measurements at 250 GeV.  The 
fit to the extended ILC program at 1~TeV includes all previous 
measurements.

\section{Results}

In Table~\ref{tab:results}, I show the results of the various fits.
The results are presented as $ 1\sigma$ confidence intervals on the
parameters $d(A)$.  Note that, for invisible Higgs decays, what is
plotted
is the square root of the branching fraction.

 In Fig.~\ref{fig:resultsLHC}, I summarize the results of the fits for 
the ultimate LHC capabilities.   The estimated errors are truly 
impressive.  If the boson at 125~GeV indeed has properties close to 
those of the Standard Model Higgs boson, the LHC experiments will 
eventually be able to demonstrate this.  However, it is unlikely that
they will reach the level of 5\% accuracy in model-independent Higgs
coupling determinations needed to distinguish the Higgs boson of 
typical new physics models from the Standard Model Higgs boson.
For this, we need another facility capable of higher precision 
Higgs boson measurements.

In Fig.~\ref{fig:resultsILC}, I summarize the results of the fits for
the various stages of the ILC.   The accuracy of the determinations
increases progressively.  The threshold measurements at 250~GeV should 
immediately attain a level of accuracy below 5\% for many of the Higgs
couplings.
However, further significant improvements are possible with Higgs
measurements at higher energies.
The substantial decrease in the accuracy of the Higgs coupling to $W$
between 250~GeV and 500~GeV results from the ability at the higher 
energy to measure the cross section for the $WW$ fusion production of
the Higgs boson in the reaction $\ee\to \nu\bar \nu h$.    This
increase in precision for $g(hWW)$ is reflected generally in the
quality of the fit.  Separation of the $gg$ and $c\bar c$ modes of
Higgs decay becomes easier at higher energies, where the Higgs is more
boosted.     The accuracies for the measurement of the rare modes
$\tau^+\tau^-$ and $\gamma\gamma$ increases progressively with higher
statistics.

High accuracy for the Higgs coupling to $t$
is realized only at the highest energies, well above the threshold for
$\ee\to t\bar t h$.  However, there is a noticeable decrease in the error
on $g(htt)$  from the LHC to the 250~GeV ILC data set, despite the
fact that there are no $t$ quark measurements at 250~GeV.  This results 
from the sharpening of other uncertainties in the fit to LHC couplings.
This is a common phenomenon in studies of the complementarity of 
LHC and ILC, one well documented in \cite{Weiglein}. 

The general conclusion is that the ILC can reach accuracies below 3\%
across
almost the complete profile of Higgs boson couplings.  The ILC will then
allow us to reach beyond the capabilities of the LHC to explore for 
small deviations from the Standard Model predictions for Higgs couplings
at the level at which these deviations are expected in models of new
physics beyond the Standard Model.

\newpage

\Acknowledgements

I thank Keisuke Fujii, Akiya Miyamoto, 
Hiroaki Ono, and Junping Tian for sharing with me
the results of the Asian full-simulation Higgs studies with the ILD detector. 
I am grateful to Yang Bai, Tim Barklow,  Kyle Cranmer,
Keisuke Fujii, Heather Logan, Tim Tait, and Jay Wacker 
for many discussions and much good advice concerning this
analysis.  I thank Gavin Salam for discussion of the
theoretical error estimates in Table 1.   I apologize to 
 these people for not accepting all of 
their suggestions. I thank Evan Friis, Sridhara Dasu, and Markus Klute for
correspondence concerning the new CMS results and Aleandro Nisati,
Michael Duehrssen, Ariel Schwartzman, and their collaborators for
correspondence concerning the new ATLAS results. 
 All responsibility for the inputs and methods used 
here is mine.   This work was supported by the US Department of
Energy under contract  DE--AC02--76SF00515.

\appendix

\section{Discussion of the new error estimates from ATLAS and CMS}

In response to the request for input to the current round of studies for
the European Strategy for Particle Physics, the ATLAS and CMS
collaborations presented new estimates of their long-term capabilites
for Higgs boson coupling measurements~\cite{ATLASESS,CMSESS}.  
 In version 2 of my
paper, I presented an update of the fit given in version 1 that takes
these new results into account.  In this Appendix, I compare these new
estimates to the values used in version 1 and explain the changes in
the values in Table 1 that I used as inputs to my fit.

Since the posting of version 2, the CMS collaboration has issued
at updated report with some additional information on their projected
capabilities
for Higgs couplings~\cite{newCMS}.  This CMS paper offers 
projections that  are very optimistic
but also completely speculative.  I give  some comments on 
this report at the end of the appendix.  The SFitter group has also 
released new estimates of the expected LHC sensitivity for 300
fb$^{-1}$ and for 3000~fb$^{-1}$~\cite{SFitter}.   The SFitter
estimates 
for 300~fb$^{-1}$ are 
 comparable to but 
slightly more pessimistic than the estimates given here, and they
expect relatively little improvement at 3000~fb$^{-1}$, due to limiting
theoretical errors.

Table~\ref{tab:compare} compares the error estimates that I used in 
version 1 to the new estimates presented by ATLAS and CMS. Most of the
numbers displaced were taken from \cite{Dthesis}. In
general, 
the experimental performance is much better for measurements similar
to those used to discover the new boson.   Where comparable estimates
exist from both ATLAS and CMS, the CMS estimates are more aggressive.
It is difficult to compare the methodologies. Because of the length
limits in the  submissions to the European Strategy study, only the
results were given, with a very brief sketch of the analysis methods.
I have consistently used the smaller of the experimental errors quoted by the two 
experiments in  Table 1 of this version 3.  

In the table, the CMS estimate for  $\tau^+\tau^-$ is broken down
into the components from the $gg$ fusion and Vector Boson fusion
 initial states~\cite{Friis}.    The
ATLAS $\tau^+\tau^-$ analysis included the VBF region only.

 For the $\sigma(WW)\cdot
BR(\gamma\gamma)$,
 $\tau^+\tau^-$ and
the $b\bar b$ measurements, I have increased  the theory errors quoted
by the experiments, appropriately, in my opinion.   For the
$\sigma(t\bar t h)\cdot
BR(\gamma\gamma)$ measurement, the ATLAS analysis is almost inclusive,
so it is claimed that the theory error should be close to that on the
total cross section for $pp\to t\bar t h$; I have used this smaller error
value in version 2.   For $\sigma(t\bar t h)\cdot
BR(b\bar b)$, I overestimated the difficulty of obtaining an accurate
QCD calculation of the total cross section, so I have reduced the
theory error from from that given version 1.  However, the poorly
understood 
uncertainties from  the boosted
Higgs technique remain.

In version 1, I included as independent measurements two pairs of
quantities that essentially measure the same observable,
$\sigma(gg)\cdot BR(ZZ)$ and $BR(\gamma\gamma)/BR(ZZ)$, and 
$\sigma(WW)\cdot BR(\tau^+\tau^-)$ and $BR(\tau^+\tau^-)/BR(ZZ)$.
 These are
indicated by the groupings in Table~\ref{tab:compare}.  In the new
estimates from ATLAS and CMS, the errors on these quantities come from
the same multivariable fit, so 
it is double-counting to include these measurements as 
independent.  For this reason, I have dropped one in each pair in
versions 2 and 3.

I have not changed the error estimates for the $b\bar b$ reactions.
In version 1, I guessed an experimental error of 25\% for each of two
reactions that were added as independent measurements.  The new CMS
valule for the error from these reactions is 18\% = 25\%/$\sqrt{2}$.

Remarkably, there is very little qualitative change in the results of
the fit from version 1 to versions 2 and 3.  This is mainly due to the
important 
role of the Higgs width to $b\bar b$, which has a
 large uncertainty  and which feeds into
all other coupling estimates through its effect on the Higgs total
width.  Overcoming this problem is a very difficult challenge for
hadron collider measurements of the Higgs properties.

The CMS report \cite{newCMS}  contains estimates of  the expected 
performance on Higgs couplings based on a 6-parameter global 
fit to expected CMS results for 300~fb$^{-1}$ and 3000~fb$^{_1}$.
CMS takes the $W$ and $Z$ 
coupling deviations to be correlated and ignores possible
 non-Standard values for
the $c\bar c$ and invisible branching fractions.   The paper reports
on two scenarios, Scenario 1, with all systematic uncertainties
unchanged, and Scenario 2, with systematic uncertainties decreasing
as $\sqrt{N}$ and theoretical errors halved.  Scenario 2 is very 
appealing.  It projects the error on the  $h\gamma\gamma$ 
coupling as 1.5\% and the error on the $hVV$ coupling as 
1.0\% after 3000~fb$^{-1}$.   Scenario 1 is already
a substantial extrapolation from current performance, since most 
of the data will be taken in an era with very high pileup. However,
it
is plausible, and is similar to the assumption adopted by ATLAS.  The 
strategy to achieve Scenario 2 is unknown.   The details of the CMS
global fit have not been provided in any public document. 

 In both scenarios, the 
large theoretical systematic error in estimating the acceptance for
the classifier that separates $pp\to W,Z + h$, $h\to b\bar b$ from 
the very similar reactions $pp \to W, Z + g$, $pp\to W,Z + Z$ seems 
to have  been ignored.  In my opinion,  it is doubtful that this acceptance
will ever be understood even to  20\%  accuracy without a breakthrough
in our QCD modelling of jet structure.  Taking into account the
formula
for 
extraction
of  $g(hb\bar b)$ from  $\sigma \cdot BR $ quantities, this
limits the expected accuracy in $g(hb\bar b)$ at about the 20\%
level. But  the
$b\bar b$ final state is the dominant component of the Higgs total
width.   All other coupling measurements depend on the
knowledge
of the Higgs total width to convert $\sigma \cdot BR $ measurements 
to coupling measurements.  This difficulty alone poisons all claims of
percent-level Higgs boson coupling measurements from the LHC.

\begin{table}
\begin{center}
\begin{tabular}{lccc} 
Observable   &   version 1 &   ATLAS    & CMS   \\ 
\hline
$\sigma(gg)\cdot BR(\gamma\gamma)$ &    0.20 $\oplus$  0.15  &   0.15
$\oplus$ 0.13 &    0.06 $\oplus$ 0.13  \\   \hline
$\sigma(WW)\cdot BR(\gamma\gamma)$ &   0.55 $\oplus$  0.10 &  0.15
$\oplus $ 0.0   &     ---   \\ \hline
$\sigma(gg)\cdot BR(ZZ)$ &                     0.21 $\oplus$  0.15
&     0.10 $\oplus$ 0.10   &      0.08 $\oplus$ 0.08 \\
$BR(\gamma\gamma)/BR(ZZ) $ &            0.21 &  0.19  &  --- \\
 \hline
$\sigma(gg)\cdot BR(WW)$ &                     ---
&     ---   &      0.09 $\oplus$ 0.11 \\
$BR(\gamma\gamma)/BR(WW) $ &            0.21 &  ---  &  --- \\
 \hline
$\sigma(gg)\cdot BR(\tau^+\tau^-)$ & --- &  --- & 0.11 $\oplus$ 
$ a$   \\ 
$\sigma(WW)\cdot BR(\tau^+\tau^-) $ &     0.22 $\oplus$  0.10  &  0.41
$\oplus$ 0.10 &  0.15 $\oplus$ $ b $  \\
$BR(\tau^+\tau^-)/BR(ZZ)$     &                  0.38 & --- & ---  \\ \hline
$\sigma(Wh)\cdot BR(b\bar b) $      &           0.25 $\oplus$ 0.20\\ 
$\sigma(Zh)\cdot BR(b\bar b) $      &           0.25 $\oplus$ 0.20\\  
$ \sigma(Vh)\cdot BR(b\bar b)$     &   ---   &    --- &    0.18
$\oplus$  0.0 \\   \hline
$\sigma(t\bar t h) \cdot BR(\gamma\gamma)$  &   0.27 $\oplus$  0.20  &
  0.42 $\oplus$ 0.10  &   --- \\
\end{tabular}
\caption{Comparison of error estimates on Higgs boson measurements
at the LHC for 1~detector and 300~fb$^{-1}$ of data
  given in version 1 of this report to those presented by ATLAS in
  \cite{ATLASESS} and by CMS in \cite{CMSESS}.  The notation is as in
  Table 1. CMS gave the total theoretical error on $\tau^+\tau^-$
  production as 0.06.}
\label{tab:compare}
\end{center}
\end{table}

\end{document}